\documentclass[final,3p,times]{elsarticle}
\usepackage[utf8]{inputenc}
\usepackage[T1]{fontenc}
\usepackage{graphicx}
\usepackage{dcolumn}
\usepackage{bm}
\usepackage{float}
\usepackage{amsmath}
\usepackage{amsfonts}
\usepackage{amssymb}
\usepackage{booktabs}
\usepackage{xcolor}

\journal{Chaos, Solitons \& Fractals}

\begin{document}
	
\begin{frontmatter}

\title{Multifractal human signals at the edge of life reveal a heart-brain anti-correlation}

\author[inst1,inst2,inst3b]{Yago Emanoel Ramos}
\ead{yago.ramos@ufpe.br}

\author[inst1]{Maria Eloá do Ó}

\author[inst3b,inst3]{Henrique Ferraz de Arruda}

\author[inst1]{Mauro Copelli} 

\author[inst4]{G. Camelo-Neto}

\author[inst1]{Pedro V. Carelli}
\ead{pedro.carelli@ufpe.br}

\affiliation[inst1]{organization={Departamento de Física, Centro de Ciência Exatas e da Natureza, Universidade Federal de Pernambuco},
            city={Recife},
            country={Brazil}}

\affiliation[inst2]{organization={Department of Theoretical Physics, Faculty of Sciences, University of Zaragoza},
            city={Zaragoza},
            country={Spain}}

\affiliation[inst3b]{organization={Institute for Biocomputation and Physics of Complex Systems (BIFI), University of Zaragoza},
            city={Zaragoza},
            country={Spain}}
            
\affiliation[inst3]{organization={ARAID Foundation},
            city={Zaragoza},
            country={Spain}}

\affiliation[inst4]{organization={Núcleo Interdisciplinar de Ciências Exatas e da Natureza, Universidade Federal de Pernambuco},
            city={Caruaru},
            country={Brazil}}

\begin{abstract}
This study investigates the terminal breakdown of human neurophysiological function through the lens of non-linear dynamics by analyzing the multifractal spectrum. Using Multifractal Detrended Fluctuation Analysis (MF-DFA), we quantify the temporal evolution of complexity in synchronized electroencephalogram (EEG) and electrocardiogram (ECG) time series from patients in the terminal stage. Our results reveal a marked divergence in multifractal spectrum width: while neural activity exhibits a collapse of multifractality toward a more constrained state, cardiac signals undergo anomalous spectral broadening, indicating increased non-linear fluctuations and dynamical instability. A negative correlation between these spectral widths suggests effective functional decoupling and the emergence of anti-correlated dynamics between neural and cardiac systems. Rather than reflecting a uniform physiological decline, this divergence is consistent with a body-to-brain breakdown in which peripheral dysfunction progressively overwhelms central regulatory processes. In a broader context, the observed opposing trends resemble patterns reported in other body-driven adaptive processes, suggesting that inverse dynamics across coupled systems may emerge when constraints originate from peripheral rather than central mechanisms. Ultimately, the dying process appears to represent an extreme form of cross-system disintegration, marked by the collapse of the hierarchical coordination that normally sustains integrated physiological function. 

\end{abstract}


\end{frontmatter}

\title{Multifractal human signals at the edge of life}
\date{}
	\section{Introduction}
    Recently, tools from complex systems have been used to study brain behavior and its control reactions in the human body\cite{Guan2025,Passaretti2025,Rajagopalan2024,Zhang2017,Bhaduri2015,Malandrone2024, Hsueh2023,Lyu2025, Qi2023,John2025}, leading to insights into brain complexity and variability on emotional behavior, motor control \cite{Passaretti2025,Ramos2025,Ramos2026}, cardiac pathologies \cite{Wong:2022}, and even the dying process \cite{Xu2023}.

    During emotionally intense states, empirical evidence suggests a reduction in the brain's functional variability, leading to more stereotyped and less flexible activation patterns. In this context, there is a decreased regulatory influence of the prefrontal cortex, a region widely associated with executive control and decision-making, alongside a relative increase in the activity of limbic structures such as the amygdala \cite{Sakaki2016}. Concurrently, cardiac signals tend to exhibit greater temporal irregularity, reflected in changes in heart rate variability, a non-invasive marker of autonomic nervous system dynamics \cite{Zhu2019}. During emotional activation, this variability may become more complex, accompanying increased physiological arousal and sympathetic dominance \cite{Sakaki2016}. Experimental studies show, for instance, that emotional intensity correlates with changes in heart rate variability (HRV) patterns as well as amygdala activation \cite{Wallentin2011}.
    
    This relationship suggests an associated brain–heart dynamics, commonly described by the neurovisceral integration model, in which cardiac variability reflects the central nervous system's capacity to regulate emotional and physiological responses \cite{Sakaki2016}. Supporting this coupling from a developmental and anatomical perspective, the heart begins to beat and function well before the brain is fully formed \cite{manner2022does}. Moreover, it possesses an intrinsic neural network, often referred to as the ``heart brain'' \cite{alshami2019pain}, that continuously sends afferent signals capable of modulating emotional and cognitive processing in the central nervous system \cite{alshami2019pain,achanta2020comprehensive}.

    Recent studies have shown that brain and cardiac signals exhibit coordinated changes in their multifractal properties, supporting the existence of a functional brain–heart interplay at the level of complex dynamics, particularly under physiologically and emotionally demanding conditions \cite{Catrambone2021}. Early studies suggest that the multifractal properties of  ECG signals are influenced by both neuroautonomic regulation and intrinsic cardiac structure, indicating that the spectra reflect a combination of neural control and electrophysiological heterogeneity \cite{Xu2005}.
    
    The multifractal spectrum is typically a broad, downward‐facing parabola-like behavior: its width $\Delta \alpha$ quantifies the range of temporal behaviors in the signal. In practical terms, a larger $\Delta \alpha$ indicates richer dynamical variability (more ``diverse'' behavior), whereas a narrow $\Delta \alpha$ implies more uniform, stereotyped dynamics \cite{Shekatkar2017,Zorick2013}. Multifractal detrended fluctuation analysis (MF-DFA) \cite{Veneziano1995,Kantelhardt2002} has become a standard tool to extract these spectra from nonstationary biomedical signals.
    
    In the cardiovascular system, this approach has revealed that healthy heartbeats tend to be more complex (i.e., have a wider $\Delta \alpha$) than those in disease. For example, Shekatkar et al.~\cite{Shekatkar2017} found that ECG attractors from healthy subjects required a wider range of scaling indices, so the multifractal width was significantly larger for healthy hearts than for pathological ones. Correspondingly, human electroencephalogram (EEG) MF-DFA reveals robust multifractality: Zorick and Mandelkern \cite{Zorick2013} showed that MF-DFA applied to waking and sleep EEG yields clear multifractal spectra, and that these spectra systematically differ across states of consciousness.
    
    Multifractal metrics have been shown to capture changes in neural dynamics associated with different levels of activity, coordination, and functional integrity \cite{francca2018fractal,stylianou2021scale,Zorick2013,racz2018multifractal}. In order to investigate how neural and cardiac dynamics are coupled under extreme physiological conditions, this study examines the relationship between the multifractal spectrum width ($\Delta \alpha$) of cardiac and neural signals during one of the most extreme states experienced by the human body:  the dying process. By using publicly available data reported by Xu et al. \cite{Xu2023}, we analyzed the multifractal spectra of cardiac and neural signals from four terminal patients during the dying process due to cardiopulmonary arrest. A baseline was defined as at least 100s preceding the withdrawal of life-sustaining support. Following this point, the patients entered the dying process while their EEG and ECG activity continued to be recorded. 
    
    This study is inherently exploratory and employs multifractal spectra computed directly from the raw signals, following the approach of Kantelhardt et al. \cite{Kantelhardt2002}, rather than HRV-based analysis. As patients progress through the dying process, heartbeats become increasingly sparse, rendering RR interval–based methods (the time interval between two consecutive R peaks in the ECG) unreliable or unfeasible.
    
\subsection{Body-to-brain dysfunction problem}

    Previous studies have demonstrated a robust functional coupling between brain and heart dynamics, a phenomenon termed functional brain-heart interplay \cite{Malandrone2024}. This framework traditionally emphasizes the role of the central autonomic network (CAN), comprising structures like the insular cortex, prefrontal cortex, and amygdala, in integrating central, peripheral, and autonomic functions through top-down neuroautonomic regulation \cite{Malandrone2024, Hsueh2023}. Consequently, cardiac dysregulation is often interpreted as a downstream effect of altered brain states, where the central nervous system processes internal and external stimuli to orchestrate autonomic responses \cite{Malandrone2024, Lyu2025}.

    However, the nature of this coupling is increasingly understood to be bidirectional, with primary bodily failures exerting causal, bottom-up influences on neural activity and affective states. For instance, optically induced tachycardia has been shown to enhance anxiety-like behavior specifically in risky contexts, a process mediated by the posterior insular cortex as a critical hub for interoceptive processing \cite{Hsueh2023}. This cardiogenic control suggests that visceral physiological signals initiate functional brain-body responses to emotional arousal \cite{Hsueh2023, Martinez2023}. In psychiatric conditions like schizophrenia, this balance is fundamentally disrupted, manifesting as weakened top-down cognitive control from the prefrontal cortex and enhanced bottom-up signaling from sensory regions, which contributes to perceptual disturbances \cite{Lyu2025}.

    In extreme physiological conditions, a functional dissociation may emerge, which may reflect a decoupling or imbalance between neural organization and physiological variability \cite{de2018dynamic}. Furthermore, the analysis of these signals can be extended through non-linear approaches such as multifractal analysis, which characterizes the temporal complexity of physiological signals through the multifractal spectrum width, providing a more sensitive measure of underlying dynamics. On the other hand, multifractal properties of neural signals are known to reflect interactions between neural and systemic (e.g., vascular) processes, rather than purely local brain activity \cite{Mukli2018}.

    Pathological phenotypes further illustrate this ``body-to-brain'' progression. Parkinson's disease exhibits a body-first phenotype, in which pathology originates in the peripheral nervous system and spreads symmetrically to the brainstem (caudal locus coeruleus) in a bottom-up fashion, leading to autonomic dysfunction before motor impairment \cite{Passaretti2025}. Similarly, in post-traumatic stress disorder, an uncoupling between the autonomic nervous system and the CAN occurs, representing a failure of integrated top-down control and a disruption of the interoceptive flow, which can be partially restored through trauma-focused psychotherapy \cite{Malandrone2024}. Such systemic disruptions are often reflected in altered multifractal dynamics; for example, psychiatric disorders such as ADHD and bipolar disorder show a significant decrease in the degree of multifractality in rs-fMRI signals across various brain regions, driven by changes in long-range temporal dependence \cite{Guan2025}.

    The dying process during cardiopulmonary arrest represents an extreme physiological state characterized by a cascading failure of interdependent organ systems \cite{Shemie2018}. Rather than being exclusively initiated or self-regulated by the brain, this process may originate from cardiac, respiratory, or neurological failure, each of which ultimately leads to global circulatory collapse and secondary brain injury. Although multisystem failure is central to this process, death is ultimately defined by the irreversible loss of brain function, which represents the final common pathway of organismal failure.
    
    Unlike recovery models such as the bimodal balance-recovery model in stroke rehabilitation, which relies on functional reorganization and structural reserve \cite{Qi2023}, the dynamics of the dying process represent terminal systemic destabilization. This collapse mirrors the spatiotemporal transitions observed in epilepsy, where initial states are marked by decreased multifractal complexity and shifts toward anti-persistent global dynamics \cite{John2025}. Such transitions provide a unique metric for quantifying network involvement in systemic failure \cite{John2025, Zhang2017}. This extreme bottom-up dysfunction involves a collapse of the interoceptive system, which typically monitors the body's homeostatic state \cite{Malandrone2024, Passaretti2025}.

    During the dying process, brain metabolism and neural function become dissociated. Experimental evidence shows that even when cellular energy metabolism is partially restored, electrophysiological activity remains severely impaired, indicating a breakdown in functional organization beyond metabolic failure \cite{nilsson1976}.
    
    Structural alterations in the brain during the dying process have also been reported, with rapid increases in tissue stiffness and reduced water diffusivity following respiratory and cardiac arrest, reflecting progressive biophysical constraints on neural tissue dynamics \cite{Bertalan2020}.

    We hypothesize that, during the dying process, brain and cardiac signals exhibit a loss of integrated regulation, reflected in signal complexity.
    
    \section{Results}

       \begin{figure}[H]
    \includegraphics[width=1\textwidth]{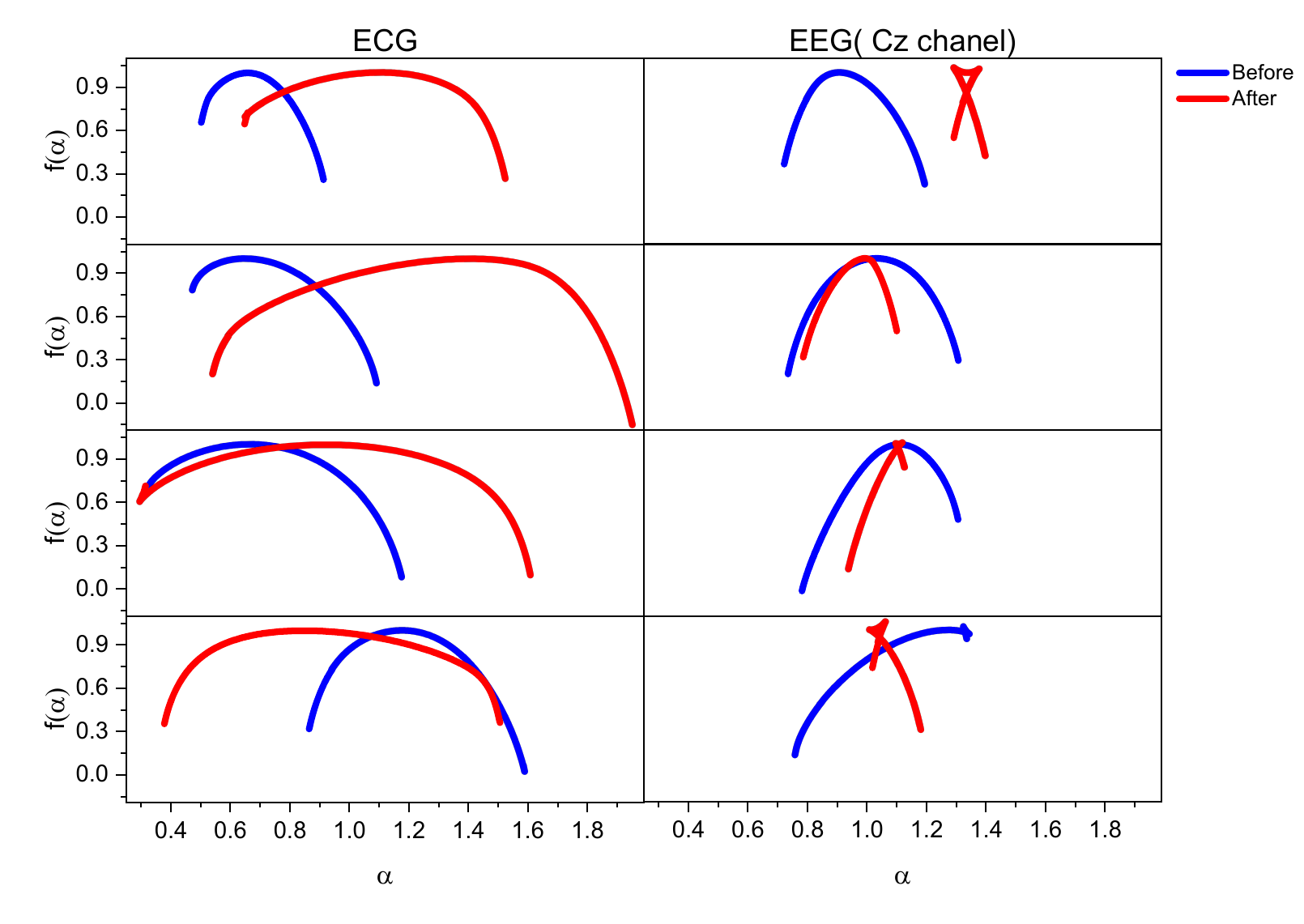}
    	\caption{Multifractal spectra derived from cardiac (first column) and neural (Cz channel, second column) data, shown for periods before and after the machines were turned off. Each line corresponds to a different patient, ordered from top to bottom as Pt1, Pt2, Pt3, and Pt4.}
    	\label{fig1}
    \end{figure}
    
    Comparing the final 100s of the baseline (before the ventilator withdrawal) with the final 100s of the patients' lives, we observed an opposite behavior in the multifractal spectrum between the cardiac and neural data: while the dying process led to a decrease in $\Delta\alpha$ in the neural data, the cardiac data exhibited a broadening of the spectrum, as shown in Fig. \ref{fig1} when comparing the ECG spectrum with that of the EEG Cz channel. A deformation of the multifractal spectrum is observed in the final stages, characterized by asymmetries and irregular tails (fish-tail effect) \cite{Lopes2024}, suggesting a reorganization of scaling properties and potential instability in the system's dynamics.
    
\begin{figure}    	\includegraphics[width=1\textwidth]{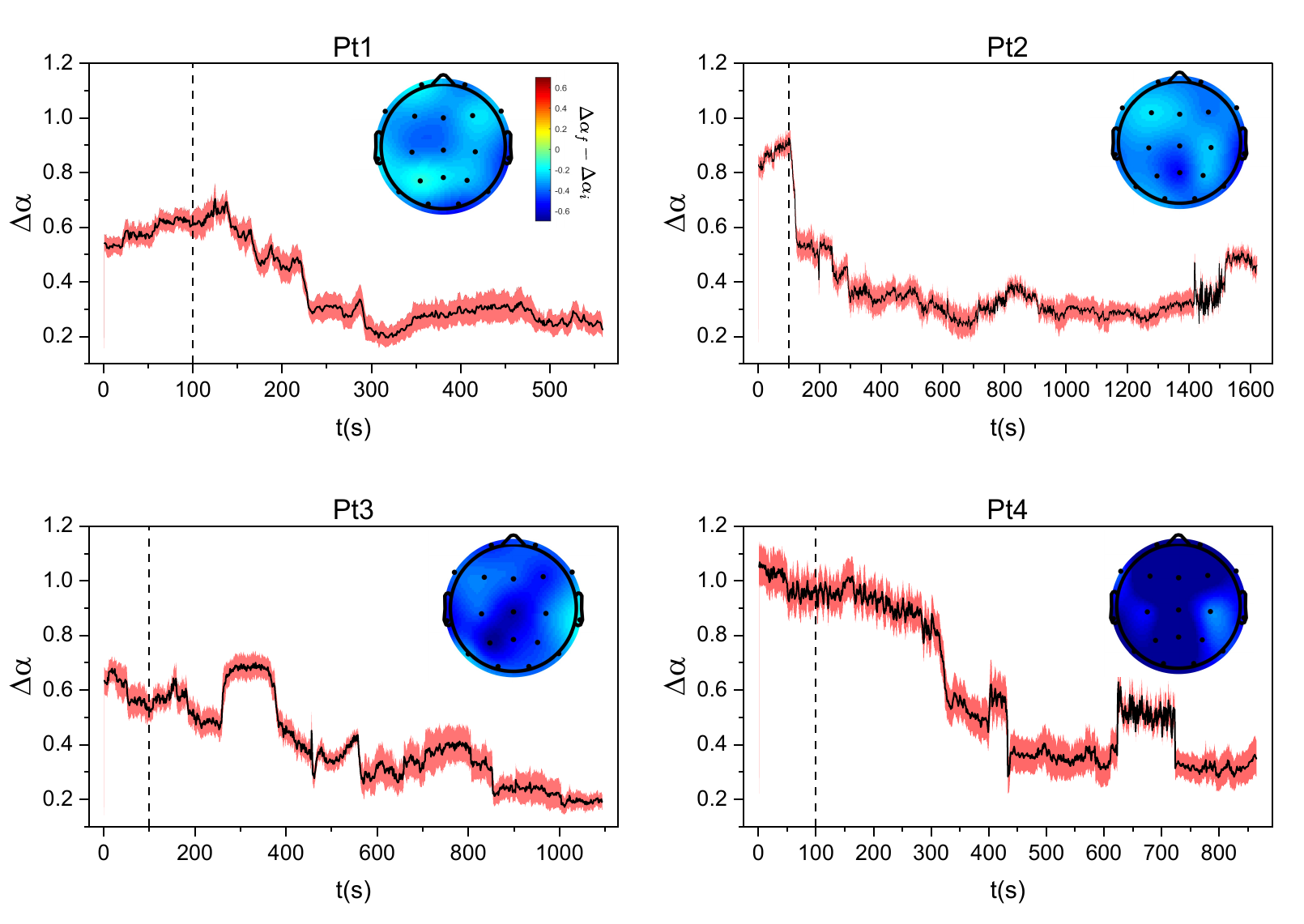}
    	\caption{Mean brain $\Delta\alpha$ between electrodes as a function of time for each patient. The shaded red region represents the $95\%$ confidence interval (CI). The dashed lines indicate the moment of ventilator withdrawal. Insets of brain topomaps are in the same colormap scale and illustrate the spatial variability of $\Delta\alpha$ up to the time of patient death.}
    	\label{fig2}
    \end{figure}

    This behavior in the neural data is consistent across all electrodes, as shown by the topomaps in Fig. \ref{fig2}, which present the spatial distribution of the difference between the sample initial and final multifractal spectrum $\Delta\alpha$ ($\Delta\alpha_{f} - \Delta\alpha_{i}$). By calculating the multifractal spectrum using a 100s sliding window with a 1s step, thereby generating a time series of multifractal spectra. We observed that the dying process is highly non-linear, lacking uniform patterns of increases and decreases in $\Delta\alpha$ (Fig. \ref{fig2}). However, there is a general trend of decreasing spectrum amplitude.
    
    Based on the results in Fig. \ref{fig2}, the variability among the electrodes is negligible compared to the temporal variability, as evidenced by the narrow confidence interval bands, which indicate that the observed trends better represent temporal rather than spatial dynamics in the brain. Therefore, we can infer that the mean accurately represents the overall brain spectrum.
   
\begin{figure}
    \includegraphics[width=1\textwidth]{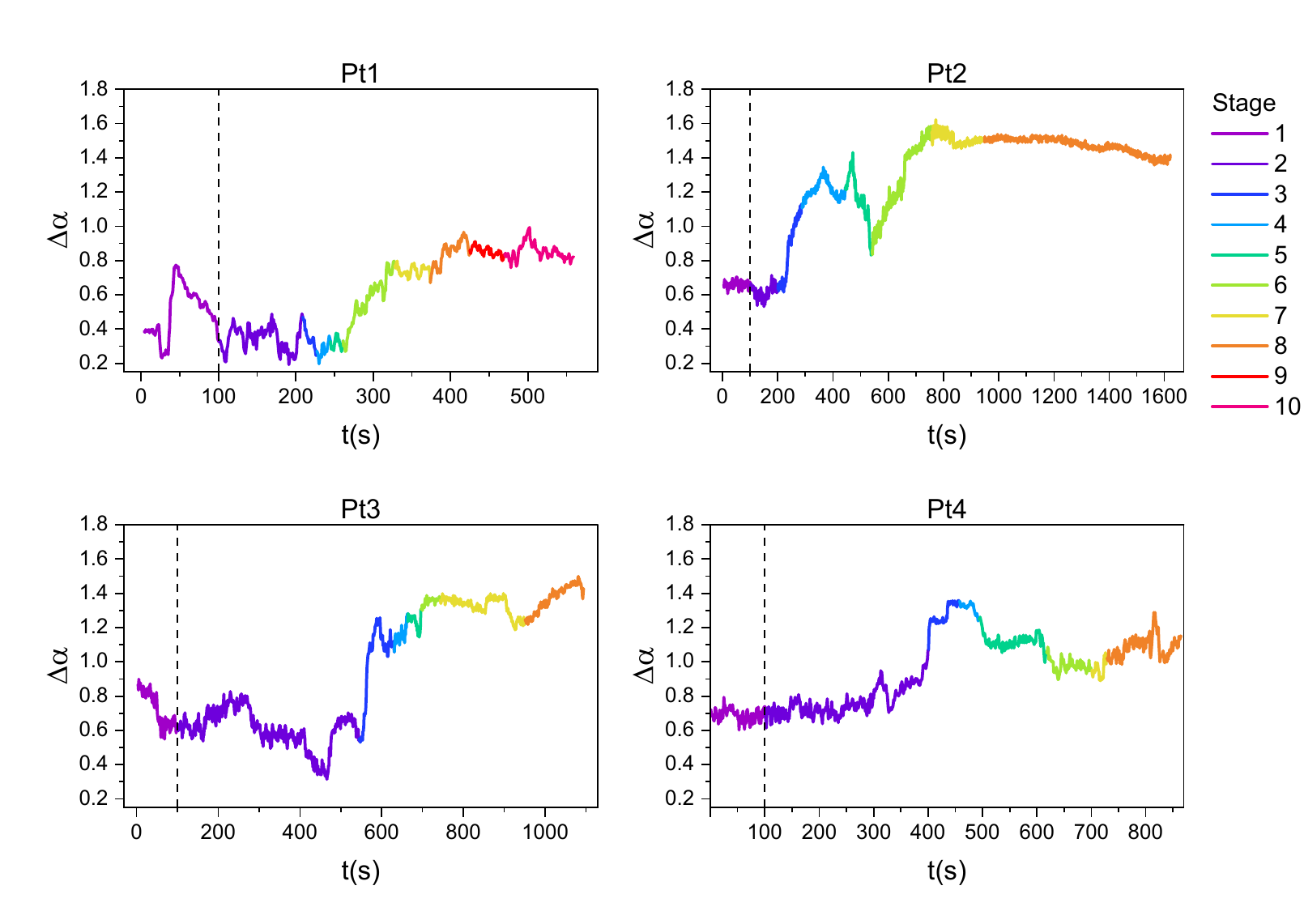}
    	\caption{Cardiac $\Delta\alpha$ as a function of time for each patient. The dashed lines represent the ventilator withdrawal moment. Colors indicate the corresponding dying stage.}
    	\label{fig3}
    \end{figure}

    The cardiac signal, however, exhibits more stereotyped behaviors, containing almost linear trends at the end of patient 2's (Pt2) signal and at the beginning of Pt4's. The signals from these ECGs were classified by Xu et al.\cite{Xu2023} into dying stages, which are shown in Fig. \ref{fig3}. The cardiac spectra demonstrate an overall trend opposite to that of the neural data, showing an increase in the $\Delta\alpha$ amplitude throughout the dying process. 
    
\begin{figure}
    	\includegraphics[width=1.0\textwidth]{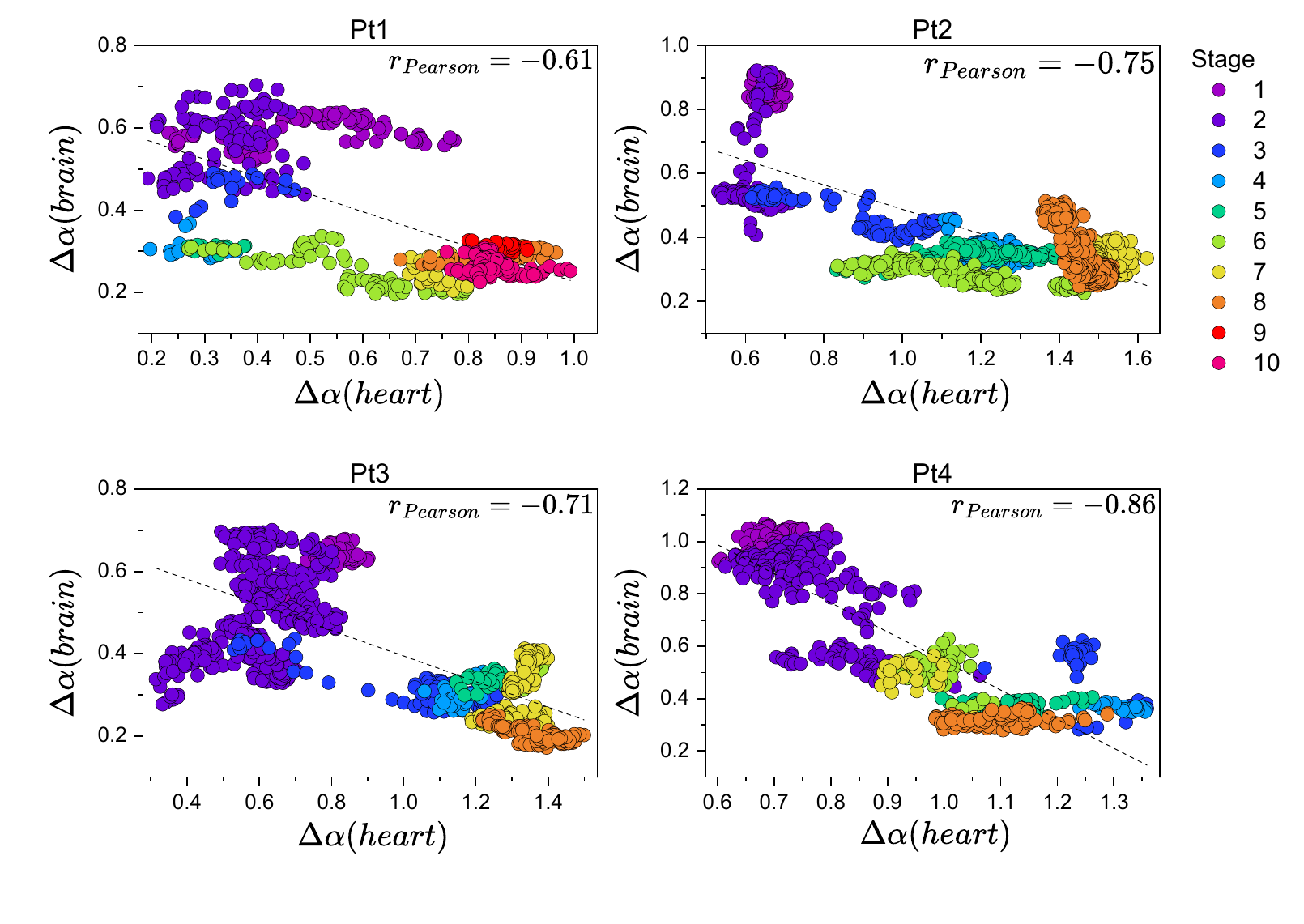}
    	\caption{Correlation between mean brain $\Delta\alpha$ and cardiac $\Delta\alpha$ for each patient. Points represent a pair of values from $\Delta\alpha$ time series from Figs. \ref{fig2} and \ref{fig3}, with colors indicating the corresponding dying stage. All p-values for the fits were less than 0.001.}
    	\label{fig4}
    \end{figure}

    In addition to the general trends observed in Figs. \ref{fig2} and \ref{fig3}, we found a negative Pearson correlation between the mean amplitude of the cerebral multifractal spectrum $\Delta\alpha(brain)$ and the cardiac spectrum $\Delta\alpha(heart)$. The Pearson correlation coefficients indicate a strong correlation for Pt2, Pt3, and Pt4 ($|r_{Pearson}| > 0.7, p<0.001$) and a moderate correlation for Pt1 ($|r_{Pearson}| > 0.5, p<0.001$). This highlights the relationship between the behavioral diversity of these two types of data during the dying process (See Fig. \ref{fig4}).
    
    Moreover, our results show a clustering of data associated with each dying stage, with the initial stages exhibiting lower $\Delta\alpha(brain)$ and $\Delta\alpha(heart)$ values. This result demonstrates that the negative linear relationship found between the $\Delta\alpha$ of these two systems acts more as a specific marker of the variation between these stages than as a biomarker for other types of conditions. 
    
	\section{Methods}
	Multifractal analysis of physiological signals depends strongly on the level of description of the system. When applied to HRV, derived from RR intervals, it primarily reflects the complexity of autonomic regulation. In this context, a broader multifractal spectrum (larger $\Delta \alpha$) is typically associated with healthy and adaptive control, while a reduction indicates loss of physiological complexity, often linked to disease or aging \cite{Ivanov1999,Goldberger2002}. These results support the interpretation of HRV multifractality as a marker of functional regulatory complexity.
	
	In contrast, multifractal analysis of the raw ECG signal captures not only timing variability but also waveform morphology, amplitude fluctuations, and intrinsic non-linear cardiac dynamics. Therefore, it reflects a more direct representation of the underlying electrophysiological system. Studies highlight that RR-based analyses may overlook relevant information contained in waveform variability, which is preserved in ECG-based approaches \cite{Shekatkar2017}. In this case, multifractality represents dynamical organization and morphological variability, rather than purely regulatory behavior.
	
	As a consequence, the interpretation of $\Delta \alpha$ differs between these signals. In HRV, higher $\Delta \alpha$ generally indicates healthy complexity, whereas in ECG, an increase in $\Delta \alpha$ may reflect either richer dynamics or pathological desynchronization. Thus, HRV and ECG multifractality should be understood as probing distinct layers of the system: control versus intrinsic dynamics. This distinction is crucial in extreme conditions, where increased multifractality in ECG may signal dynamical breakdown, while decreased multifractality in other systems reflects a loss of functional organization.
	
    \subsection{Data Acquisition and Preprocessing}

	Nineteen continuous EEG channels from the international 10/20 system and one heart recording were imported for each participant; sequential recording sessions were temporally concatenated. The data, sampled at a frequency of 512 Hz, were segmented using a sliding window approach to capture dynamic temporal variations. No bandpass filtering was applied.
    
    A window size of 100 seconds was applied with a 1-second sliding step, resulting in a 99\% overlap between consecutive epochs. 
    
    The original data have 120s of baseline (state 1) for patient Pt1 and 100s for Pt2, Pt3 and Pt4. In order to unify the comparisons, we considered only the last 100s of the Pt1 baseline. A notch filter was applied to remove the 60-Hz artifact from EEG signals.
    
	\subsection{Dying stages}
	In the study by Xu et al. \cite{Xu2023}, cardiac dying stages were operationally defined based on combined EEG and ECG dynamics. The baseline stage (1) corresponds to the period under full life support prior to ventilator withdrawal. State 2 spans from ventilator removal to the onset of acute EEG suppression. Subsequent stages (3 onward) are delineated according to progressive alterations in cardiac activity, characterized using ECG-derived features via the electrocardiomatrix method introduced by Li et al.\cite{li2015electrocardiomatrix}, thereby capturing the temporal evolution of neuro-cardiac deterioration during the dying process.

	\subsection{Multifractal Detrended Fluctuation Analysis (MF-DFA)}

    To characterize the non-linear dynamics and multifractal properties of the signals, the MF-DFA method was applied to each channel within every windowed segment, following the framework proposed by Kantelhardt et al.~\cite{Kantelhardt2002}.

Given a  time series $x_k$ of length $N$, the integrated profile was computed as:
\begin{equation}
y(k) = \sum_{i=1}^{k} (x_i - \bar{x}),
\end{equation}
as originally introduced in the context of detrended fluctuation analysis by Peng et al.~\cite{peng1994mosaic}.

The integrated profile was then divided into $N_s = \lfloor N/s \rfloor$ non-overlapping segments of equal length $s$. In this study, 25 logarithmically spaced scales were considered, ranging from $s_{\min} = \max(16, N/1024)$ to $s_{\max} = N/4$. Within each segment $\nu$, a first-order polynomial fit was used to remove local trends, and the variance of the detrended signal was computed as $F^2(s,\nu)$.

The $q$-order fluctuation function was then defined as:
\begin{equation}
F_q(s) = \left( \frac{1}{N_s} \sum_{\nu=1}^{N_s} \left[ F^2(s,\nu) \right]^{q/2} \right)^{1/q},
\end{equation}
for $q \neq 0$. For $q=0$, a logarithmic averaging procedure is required:
\begin{equation}
  F_0(s) \equiv \exp \left\{ \frac{1}{2N_s} \sum_{\nu=1}^{2N_s} \ln \left[ F^2(\nu,s) \right] \right\}.
  \label{eq:fluctuation_0}
\end{equation}
This formulation generalizes the standard DFA by introducing a dependence on the parameter $q$, allowing the characterization of different fluctuation magnitudes~\cite{Kantelhardt2002}.

In this work, $F_q(s)$ was evaluated for 1200 values of $q$ uniformly distributed in the range $[-10, 10]$, enabling a detailed description of both small and large fluctuations. The generalized Hurst exponent $H(q)$ was estimated as the slope of the linear regression between $\log F_q(s)$ and $\log s$.
\begin{equation}
  F_q(s) \sim s^{H(q)}.
  \label{eq:scaling_law}
\end{equation}

The multifractal scaling exponent $\tau(q)$ was then obtained as follows:
\begin{equation}
\tau(q) = qH(q) - 1,
\end{equation}
which is directly related to the classical multifractal formalism.

The singularity strength $\alpha$ and the multifractal spectrum $f(\alpha)$ were computed via the Legendre transform:
\begin{equation}
\alpha = \frac{d\tau(q)}{dq}, \quad f(\alpha) = q\alpha - \tau(q),
\end{equation}
providing a geometrical interpretation of the distribution of singularities in the signal.

Finally, the width of the multifractal spectrum:
\begin{equation}
\Delta \alpha = \alpha_{\max} - \alpha_{\min},
\end{equation}
was extracted as the main quantitative descriptor of multifractality, with larger values indicating a higher degree of complexity and heterogeneity in the temporal dynamics~\cite{ihlen2012introduction}.
    \subsection{Sliding windows and methodological limitations}

    The choice of a 100 s window is justified by the limited length of the baseline data. To ensure that our results are not biased by the limited baseline length restriction, we set the window length to the maximum possible value, allowing a consistent comparison of the evolution across the dying stages.
    
    While specific physiological stages are precisely timestamped in the original recordings, the application of a sliding window approach for MF-DFA introduces inherent constraints on temporal resolution and boundary definitions. 
    
    The primary limitation arises from the window length $W = 100s$. For a raw time series of total duration $T_{total}$, the resulting multifractal spectrum series $\Delta\alpha(t)$ has a reduced effective duration, $T_{eff}$, defined by:

  \begin{equation}
      T_{eff} = T_{total} - W + \delta t,
  \end{equation}
   where $\delta t$ represents the sliding step (1s). Consequently, the first data point in our analysis represents the integral complexity of the initial 100s baseline period (state 1). As the window slides, the transition between physiological stages is smoothed rather than discrete. The boundary where the baseline is completely excluded from the calculation—indicated by the dashed lines in Figs. \ref{fig2} and \ref{fig3}, occurs at:

    \begin{equation}
        t_{limit} = t_{start} + W.
    \end{equation}
    
    This temporal averaging leads to an unavoidable ``edge effect'' at the end of the recordings. Specifically, terminal stages that occur within the final $W$ seconds of the data collection are absorbed into the final window calculation, preventing them from being resolved as independent dynamical stages. This accounts for the absence of the final recorded stages in our analyzed series, specifically stage 11 for Pt1 and stage 9 for Pt4. 

    Despite these limitations, which are inherent to the non-stationary nature of physiological signals, the resulting $\Delta\alpha$ trajectories demonstrate robust stability. The emergence of clearly separable and consistent clusters in Fig. \ref{fig4} suggests that the windowed approach successfully captures the underlying dynamical transitions of the heart-brain system, even while operating under these temporal constraints.
    
    \section{Discussion and Conclusion}
    
   When considering these opposing trends between bain and heart electrical activity, an interesting parallel emerges with previous studies examining processes driven by peripheral constraints rather than by changes in central control. For example, investigations of writing with the non-dominant hand \cite{Ramos2025} and collective dance improvisation under coordinated group constraints \cite{Ramos2026} revealed inverse patterns between brain and movement dynamics. Although these contexts differ substantially from the dying process, they suggest that opposing trends across coupled physiological or behavioral systems may arise when novel demands originate from the body or from peripheral interactions rather than from the central nervous system itself.

   Viewed alongside these earlier findings, the present results extend a pattern previously observed in human movement to the context of cardiorespiratory system failure. A common feature across these scenarios is that the primary challenge emerges from altered bodily conditions. Individuals already know how to write, yet writing with the non-dominant hand imposes new motor constraints. Dancers already know how to improvise, yet collective coordination under specific group dynamics introduces additional demands. Likewise, breathing is normally an automatic and highly regulated process, but during dying, the body progressively loses the capacity to sustain it. In each case, adaptation is driven less by the acquisition of new central control mechanisms and more by the need to respond to increasing constraints arising from peripheral systems. This perspective is consistent with the hypothesis that the dying process represents a body-to-brain breakdown, in which the progressive loss of physiological integration originates outside the central nervous system and ultimately disrupts organism-wide coordination.
    
    These findings support the hypothesis of a body-to-brain breakdown, in which systemic failure is initiated by peripheral organs rather than through top-down neural control. As this process unfolds, cardiac signals exhibit a broadening of behavioral patterns, including pathological and unregulated components, while the brain progressively loses oxygen supply, functional complexity, and regulatory capacity. Unlike most pathological conditions, in which the brain retains some degree of influence over bodily processes, the dying process appears to represent a transition toward the loss of integrated physiological control.

    More broadly, the observed opposing trends between coupled physiological systems may reflect a general organizational principle that emerges when adaptive demands originate from peripheral processes rather than from central regulation. Similar patterns have been reported in contexts such as motor adaptation and collective movement, where established central functions remain available but must respond to novel bodily or interactional constraints. From this perspective, the dying process may represent an extreme manifestation of the same phenomenon, in which escalating peripheral failure progressively overwhelms the organism's capacity for integrated regulation. Ultimately, these results suggest that the end of life is characterized by the progressive collapse of the cross-system coordination that normally sustains physiological control, or more simply, by the progressive loss of homeostasis.

\section{Acknowledgments}

We thank the Brazilian Federal Agency for Support and Evaluation of Graduate Education (CAPES) for the Yago Emanoel Ramos grant (88887.203818/2025–00). Maria Eloá and G. Camelo-Neto would like to thank the Fundação de Amparo à Ciência e Tecnologia do Estado de Pernambuco (FACEPE) for the financial support and the fellowship granted during this research (IBPG-2329-1.05/25, APQ-1636-1.06/24 and BIC-2191-1.05/25). G. Camelo Neto also acknowledge FACEPE, CAPES, Conselho Nacional de Desenvolvimento Científico e Tecnológico (CNPq), and Financiadora de Estudos e Projetos (FINEP) for providing partial financial support under grants APQ-1056-1.06/22, 23076.067168/2021-40, 23076.090159/2021-83, 444500/2024-3, and 308703/2022-7. We also thank FADE/UFPE grant 64/2024. This research is supported by INCT-NeuroComp (CNPq Grant 408389/2024-9). We thank Arthur Vinícius de Oliveira, for valuable discussions and critical review of the text.

\bibliographystyle{elsarticle-num}
\bibliography{references}

\end{document}